\documentclass[10pt,twocolumn,letterpaper]{article}
\usepackage{usenix2019}

\newif\ifcameraready
\camerareadyfalse

\usepackage{lmodern}
\usepackage{amssymb,amsmath,amsfonts}
\usepackage[T1]{fontenc}
\usepackage[utf8]{inputenc}

\usepackage{upquote}

\usepackage[sort]{cite}

\usepackage{microtype}
\UseMicrotypeSet[protrusion]{basicmath} %

\usepackage{hyperref}
\usepackage{graphicx,grffile}

\usepackage{xspace}
\usepackage{times}
\usepackage{multirow}
\usepackage{xspace}
\usepackage{tabularx}
\usepackage{ragged2e}
\usepackage{booktabs}
\usepackage{paralist}
\usepackage[american]{babel}
\usepackage[shortlabels]{enumitem}
\usepackage{color,wrapfig}
\usepackage{balance}
\usepackage{enumitem}
\usepackage{epstopdf}
\usepackage{booktabs}
\usepackage{url}
\usepackage{pifont}
\usepackage{comment}
\usepackage{mathtools}
\usepackage[normalem]{ulem}
\usepackage{xkeyval}
\usepackage[export]{adjustbox}
\usepackage{subfig}
\usepackage{amsmath}

\usepackage{xcolor}
\usepackage{beramono}
\usepackage{listings}

\definecolor{mGreen}{rgb}{0,0.6,0}
\definecolor{mGray}{rgb}{0.5,0.5,0.5}
\definecolor{mPurple}{rgb}{0.58,0,0.82}
\definecolor{backgroundColour}{rgb}{0.95,0.95,0.92}

\usepackage[ruled,noend]{algorithm2e}
\LinesNumbered
\usepackage[margin=2pt,font=small,labelfont=bf]{caption}
\usepackage{etoolbox}
\usepackage{ifthen}
\usepackage{sepfootnotes}
\usepackage[compact]{titlesec}

\usepackage[disable]{todonotes}

\usepackage{tikz}
\usetikzlibrary{shapes.geometric}
\usetikzlibrary{shapes.multipart}
\usetikzlibrary{shapes.misc}
\usetikzlibrary{arrows}
\usetikzlibrary{arrows.meta}
\usepackage{pgfplots}
\usepackage{pgfplotstable}
\usepgfplotslibrary{fillbetween}
\usetikzlibrary{pgfplots.groupplots}
\pgfplotsset{compat=1.15}

\PassOptionsToPackage{usenames,dvipsnames}{color} %
\hypersetup{unicode=true,
            colorlinks=true,
            linkcolor=blue,
            citecolor=blue,
            anchorcolor=blue,
            urlcolor=blue,
            breaklinks=true}
\makeatletter
\def\maxwidth{\ifdim\Gin@nat@width>\linewidth\linewidth\else\Gin@nat@width\fi}
\def\maxheight{\ifdim\Gin@nat@height>\textheight\textheight\else\Gin@nat@height\fi}
\makeatother
\setkeys{Gin}{width=\maxwidth,height=\maxheight,keepaspectratio}
\makeatletter
\g@addto@macro{\UrlBreaks}{\UrlOrds}
\makeatother

\newcommand\paraspace{\vspace*{0.5ex}}
\providecommand\parab[1]{\paraspace\noindent\textbf{#1}}
\providecommand\parae[1]{\paraspace\noindent\textit{#1}}

\apptocmd\normalsize{%
\abovedisplayskip=2pt
\abovedisplayshortskip=2pt
\belowdisplayskip=2pt
\belowdisplayshortskip=2pt
}{}{}

\lstdefinestyle{CStyle}{
    backgroundcolor=\color{backgroundColour},
    commentstyle=\color{mGray},
    keywordstyle=\color{magenta},
    numberstyle=\tiny\color{mGray},
    stringstyle=\color{mPurple},
    basicstyle=\footnotesize\ttfamily\SetTracking{encoding=*}{-60}\lsstyle,
    breakatwhitespace=false,
    breaklines=true,
    captionpos=b,
    keepspaces=true,
    numbers=left,
    numbersep=5pt,
    showspaces=false,
    showstringspaces=false,
    showtabs=true,
    tabsize=2,
    language=C
}

\lstdefinestyle{PyStyle}{
    backgroundcolor=\color{backgroundColour},
    commentstyle=\color{mGray},
    keywordstyle=\color{magenta},
    numberstyle=\tiny\color{mGray},
    stringstyle=\color{mPurple},
    basicstyle=\footnotesize\ttfamily\SetTracking{encoding=*}{-60}\lsstyle,
    breakatwhitespace=false,
    breaklines=true,
    captionpos=b,
    keepspaces=true,
    numbers=left,
    numbersep=5pt,
    showspaces=false,
    showstringspaces=false,
    showtabs=true,
    tabsize=2,
    language=Python
}

\newcommand{\sysname}{Galleon\xspace}
\newcommand{\system}{Galleon\xspace}

\newcommand{\ie}{\emph{i.e.,}\xspace}
\newcommand{\eg}{\emph{e.g.,}\xspace}

\newcommand{\secref}[1]{\S\ref{#1}}
\newcommand{\figref}[1]{Figure~\ref{#1}}

\ifcameraready

\fi

\begin{document}

\title{\LARGE\system: Reshaping the Square Peg of NFV} %

\author{
{\rm Jianfeng Wang}\\
University of Southern\\ California\\
\and
\rm{Tamás Lévai}\\
Budapest University of\\Technology and Economics\\
\and
\rm{Zhuojin Li}\\
University of Southern\\ California\\
\and
\rm{Marcos A. M. Vieira}\\
Universidade Federal de \quad\\Minas Gerais\\
\and
\rm{Ramesh Govindan}\\
University of Southern\\ California\\
\and
\quad \rm{Barath Raghavan}\\
\quad University of Southern\\ California\\
}

\maketitle

\begin{abstract}
Software is often used for Network Functions (NFs)---such as firewalls, NAT, deep packet inspection, and encryption---that are applied to traffic in the network. The community has hoped that NFV would enable rapid development of new NFs and leverage commodity computing infrastructure. However, the challenge for researchers and operators has been to align the square peg of high-speed packet processing with the round hole of cloud computing infrastructures and abstractions, all while delivering performance, scalability, and isolation.

Past work has led to the belief that NFV requires custom approaches that deviate from today's abstractions. To the contrary, we show that we can achieve performance, scalability, and isolation in NFV judiciously using mechanisms and abstractions of FaaS, the Linux kernel, NIC hardware, and OpenFlow switches. As such, with our system \system, NFV can be practically-deployable today in conventional cloud environments while delivering up to double the performance per core compared to the state of the art. 
\end{abstract}

\section{Introduction}
\label{sec:intro}

Software is often used for Network Functions (NFs)---such as firewalls, NAT, deep packet inspection, and encryption---that are applied to traffic in the network. For the last eight years, the hope has been that NF Virtualization (NFV) would enable rapid development of new NFs by multiple software vendors and leverage the power and economics of commodity computing infrastructure. The challenge for the systems community and for network operators has been to align the square peg of high-speed packet processing with the round hole of cloud computing infrastructures and abstractions.

To this end, many NFV solutions have been proposed to provide performance, scalability, and isolation, each with novel mechanisms and inherent tradeoffs.  Some have explored how to integrate hardware and software to achieve high performance and scalability~\cite{Metron2018, YenWang2020Lemur, UNO2017} but ignore NF isolation; others have solved isolation but require programming language support~\cite{NetBricks2016}; still others have used appropriate cloud abstractions like FaaS but have not addressed scalability~\cite{ServelessNF}. 

The research community has believed that NFV is different enough that it requires novel, custom approaches that deviate from today's norms. To the contrary, we show that NFV may be a round peg after all---that we can achieve performance, scalability, and isolation using the standard mechanisms and abstractions of today's cloud computing environments---building upon Function-as-a-Service (FaaS), the Linux kernel, standard NIC hardware, and OpenFlow switches. As such, with \system, NFV can be deployable today in conventional cloud environments.

Performance, scalability, and isolation are key for NFV to be production ready.  NFV workloads involve the deployment of chains of one or more NFs from different vendors, in a multi-tenant environment; for deployment and management ease, NFV needs to conform to the infrastructure rather than the other way around.
Offloading some NF processing to specialized hardware, for example, entails operational complexity.  NF vendors implement NFs as they see fit and make them available as containers or VMs, giving operators little language or hardware choice. Indeed, customization of the infrastructure undercuts a key focus of NFV: to eliminate the operational headaches of the hardware-based middlebox era.

NFV, as an important workload for the edge, is often involved in rack-scale deployments at telco central offices or ISP PoPs. Further, 
with limited commodity hardware, such a platform often puts an efficiency requirement for scaling up and down both NFV chains and other edge applications quickly. Further, it prefers not to split the infrastructure into smaller subsets that serve one type of dedicated applications within each subset. This not only increases the management overhead of the platform, but also prevents it from multiplexing the infrastructure among all deployed applications, which further reduces the system efficiency.

We posit that NFV must embrace a dirty-slate approach.
In this paper, we ask: is it possible to achieve performance, scalability, and isolation in NFV on general-purpose infrastructure? We believe that three key enabling pieces are missing from today's NFV literature: first, the ability of deploying third-party NFs with an efficient isolation mechanism without losing generality; second, the ability to run NF chains with an efficient yet standard scaling mechanism; third, the ability to execute NF chains that coexist with other edge applications while meeting SLO targets.

This paper makes the following contributions:\\[0.5ex]
\parab{FaaS compatibility.} \sysname minimally extends FaaS abstractions to enable them to support NFV workloads. As in existing FaaS implementations, each NF is a separate container. NFs are, however, triggered on packet arrival, not on application level request arrival. \sysname builds upon OpenFaaS~\cite{openfaas} and leverages Kubernetes so that this optimized FaaS infrastructure maintains its compatibility with other applications while meeting the needs of NFV.\\[0.5ex]
\parab{Performance-aware scheduling.} \sysname dedicates cores to NF chains and uses kernel bypass to deliver packets to NFs. It uses standard OS interfaces to cooperatively schedule NFs in a chain, to mimic the \textit{run-to-completion}~\cite{Metron2018} strategy that has proven to be essential for high NFV performance. Run-to-completion processes a batch of packets; \sysname selects batch sizes that satisfy SLOs while minimizing context switch overhead.\\[0.5ex]
\parab{Efficient scaling with SLO adherence.} In response to changes in traffic, \sysname auto-scales NF chain instances in a manner that minimizes core usage while preserving latency SLOs. This flexibility allows tenants to trade-off latency for lower cost, a capability present in other NFV systems~\cite{tootoonchian2018resq}.\\[0.5ex]
\parab{High-performance spatiotemporal packet isolation.} \sysname's use of containerized NFs, together with NIC virtualization, ensures that an NF chain can only see its own traffic. \sysname also ensures a stronger form of packet isolation: an NF in a chain can process a packet only after its predecessor NF. It does this by spatially isolating the first NF from the others in the chain using a packet copy. Subsequent NFs can process packets in zero-copy fashion, with temporal isolation enforced by scheduling. This approach is general and transparent to NF implementations, and requires neither language support nor high-overhead mechanisms.\\[0.5ex]
We find \sysname achieves up to 1.93$\times$ the per-core throughput of state-of-the-art NFV systems~\cite{NetBricks2016,edgeOS2020} that use alternative isolation mechanisms. Under highly dynamic loads, \sysname achieves zero packet losses and is able to satisfy tail latency SLOs. Compared to a highly-optimized NFV system that does not provide packet isolation and is not designed to satisfy latency SLOs (but is designed to minimize latency)~\cite{Metron2018}, \sysname uses slightly more CPU cores (15\%) while achieving isolation \textit{and} satisfying latency SLOs.

\section{Background}
\label{sec:problem-definition}

\subsection{NFV's promise}
\label{s:nfv-workloads}

NFV's promise was to replace bulky hardware middleboxes with virtualized, easier-to-manage software-based NFs. NF functionality varies widely, from simple (\eg VLAN tunneling) to complex (\eg traffic inference). NFs are chained together to process packets in sequence to meet a network operator's needs; a canonical chain employs a firewall, deep packet inspection, and encrypted tunneling.

NFV workloads have demanding performance requirements, as they must forward packets at hundreds of Gbps while meeting stringent per-chain latency SLOs. It is also important to achieve high resource efficiency for deploying such workloads so that the NFV service providers can serve traffic with less resources, and further reduce the cost of NFV.

\subsection{\sysname's Goal}
\label{s:faas-nfv-goals}

While NFV initially promised to adopt cloud computing abstractions, the move from specialized hardware to commodity servers running canonical software stacks has been in name only.  Indeed, the \textit{state of the art in NFV employs custom interfaces, run-times and control planes}~\cite{E22015, Metron2018,NetBricks2016,edgeOS2020,kulkarni2017nfvnice}, breaks abstraction boundaries in the name of performance~\cite{E22015, Metron2018}, places a greater burden upon the programmer to ensure key properties such as isolation~\cite{NetBricks2016}, and leverages specialized hardware to achieve high performance~\cite{YenWang2020Lemur}. These \textit{customized NFV platforms} have been absolutely essential for the intellectual advancement of the area, but they are insufficient to enable general, widespread NFV offerings in practice.

Our goal with \sysname is to align NFV with ubiquitous cloud abstractions
without sacrificing performance, scalability, or isolation. In particular, our
aim is to build upon the abstractions and mechanisms available with
Function-as-a-Service (FaaS) platforms. FaaS closely aligns
with NFV's needs: to execute a modular piece of code (\eg an NF) over 
discrete units of data (\eg packets). Recent prior work ServerlessNF~\cite{SNF,ServelessNF} aims to achieve these goals on unmodified FaaS platforms. In contrast, \sysname proposes minimal additional abstractions for packet processing that enable it to provide performance and feature sets comparable to customized NF processing platforms (\eg Metron~\cite{Metron2018}).

FaaS is a good candidate for deploying NFV. NF developers can focus on developing new NFs and leave the orchestration and management overhead to FaaS providers. With FaaS’s fine-grained billing, NF developers can also benefit from a cost-efficient deployment when serving dynamic traffic.  Developers write and upload either a piece of code or an executable, which is then executed in response to incoming requests. FaaS providers hide the underlying infrastructure from developers, and handle provisioning, executing, scheduling, scaling, and resource accounting of user-defined code. FaaS comes with built-in scalability, and pursues extreme efficiency such that deployments do not pay for resources that they do not use.

While these benefits sound plausible, today’s FaaS platforms cannot serve NFV workloads with demanding performance requirements. In fact, most existing FaaS platforms do not offer any performance guarantees for underlying applications. They employ non-deterministic scheduling on top of shared resources. This makes them mostly useful for latency-insensitive applications. While executing modular functions, FaaS platforms lack an efficient inter-function communication mechanism and often rely on third-party cloud storage services for inter-function communications. This introduces new sources of latency, and can be extremely cost-inefficient when almost all packets must be transmitted among NFs.

\sysname's aims to provide a FaaS platform suitable for deploying NF chains in a multi-tenant environment. As such, it must satisfy four requirements. The first of these is unique to \sysname, while the remaining requirements ensure that \sysname is comparable to custom NFV platforms.

\parab{Abstractions and Interfaces.} Current FaaS systems provide REST-ful interfaces inappropriate for packet processing~\cite{ServelessNF}. \sysname must 1) introduce minimal new abstractions for NFV while maintaining performance comparable to standalone platforms like Metron~\cite{Metron2018}, NetBricks~\cite{NetBricks2016}, or NFVnice~\cite{kulkarni2017nfvnice}, and 2) must do so without impacting the core FaaS abstractions that permit application-level functions. In developing these abstractions, the following challenges arise: How do NF developers interact with the system to develop new NFs? What information must be collected via this interface to specify the requirement of deploying NF chains? How does this interface co-exist with existing cloud interfaces?

\parab{Performance-aware scheduling.} Each incoming packet triggers the execution of an NF chain. Thus, ingress traffic needs to be distributed to server CPU cores that are tasked with executing the target NF chain. There must be mechanisms for 1) routing packets so that they are processed by the right chain, 2) scheduling NFs to process incoming packets on underlying hardware resources (i.e., CPU cores), and 3) meeting per-chain SLO requirements. Prior approaches break abstraction boundaries and assume programmer involvement in performance tuning. \sysname cannot do this; fortunately, however, it can use techniques (\eg kernel bypass, virtualized NICs) developed for low-latency cloud communication.

\parab{Dynamic scaling.} Dynamic scaling needs to quickly adapt to load changes and provision server resources. Such scaling is already built into FaaS offerings, but \sysname needs to define fine-grained low-overhead monitoring techniques, as well as policies for scaling NF chains. Unlike NFV platforms, \sysname must minimize core usage while scaling (so cluster resources can be used by other non-NFV FaaS workloads).

\parab{Isolation.} Many NFV platforms provide \textit{packet isolation} that \sysname must also support: an NF should not be able to access a packet until its predecessor NF in the chain has finished processing the packet, and NFs in an NF chain should not be able to see packets destined to another chain. This requirement arises in multi-tenant settings where each chain may consist of NFs from multiple vendors, and each chain may be responsible for processing a specific customer's traffic.

In \secref{sec:design}, \secref{sec:cooperative-sched}, and \secref{sec:faas-controller}, we show how \sysname achieves these.

\section{\sysname Overview}
\label{sec:design}

\begin{figure}[t]
\begin{center}
\subfloat[\sysname's control plane.]{
\resizebox{\columnwidth}{!}{%
    \includegraphics[width=0.45\textwidth]{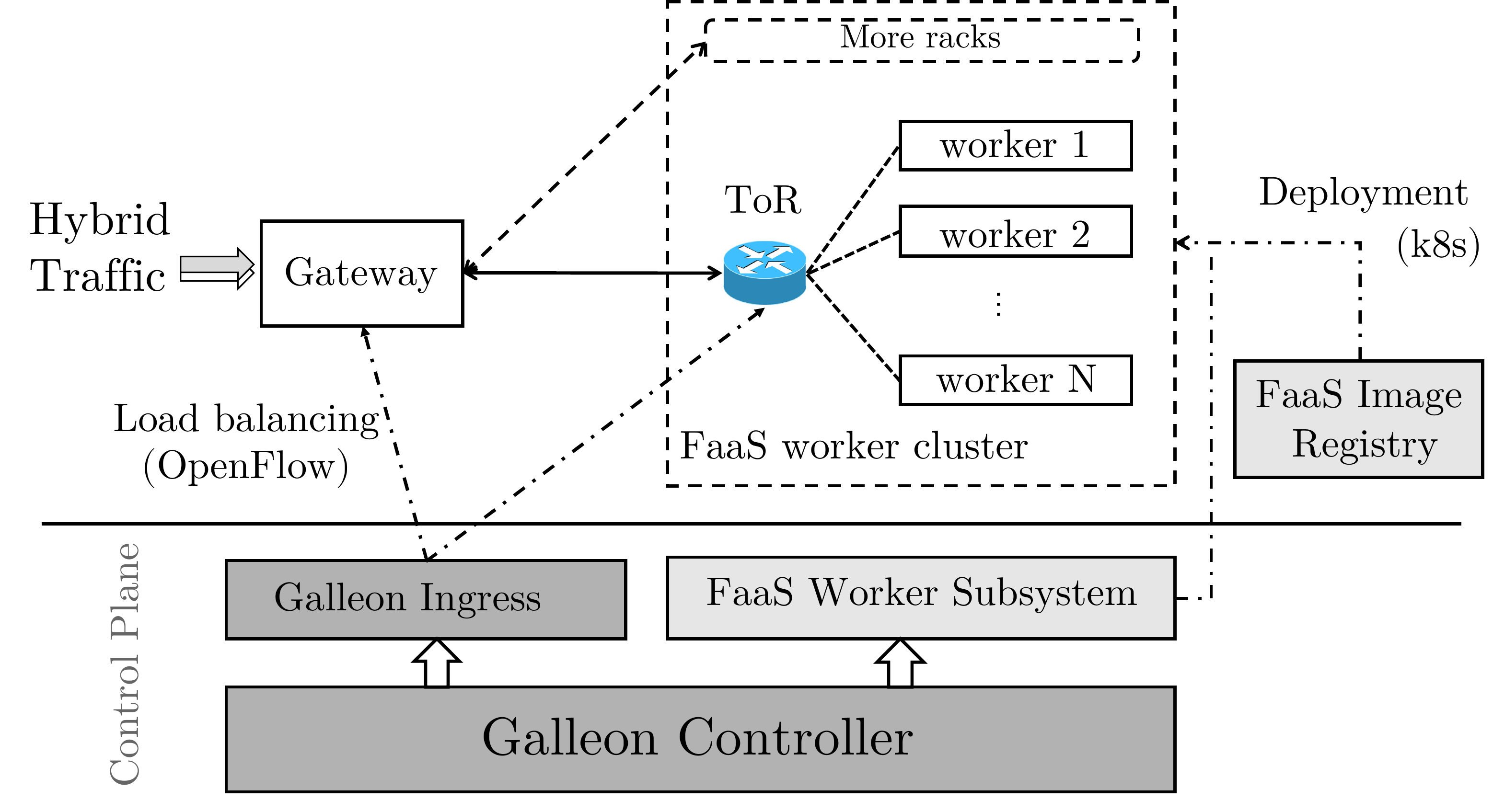}
}
}\\
\subfloat[\sysname worker.]{
\resizebox{\columnwidth}{!}{%
    \includegraphics[width=0.09\textwidth]{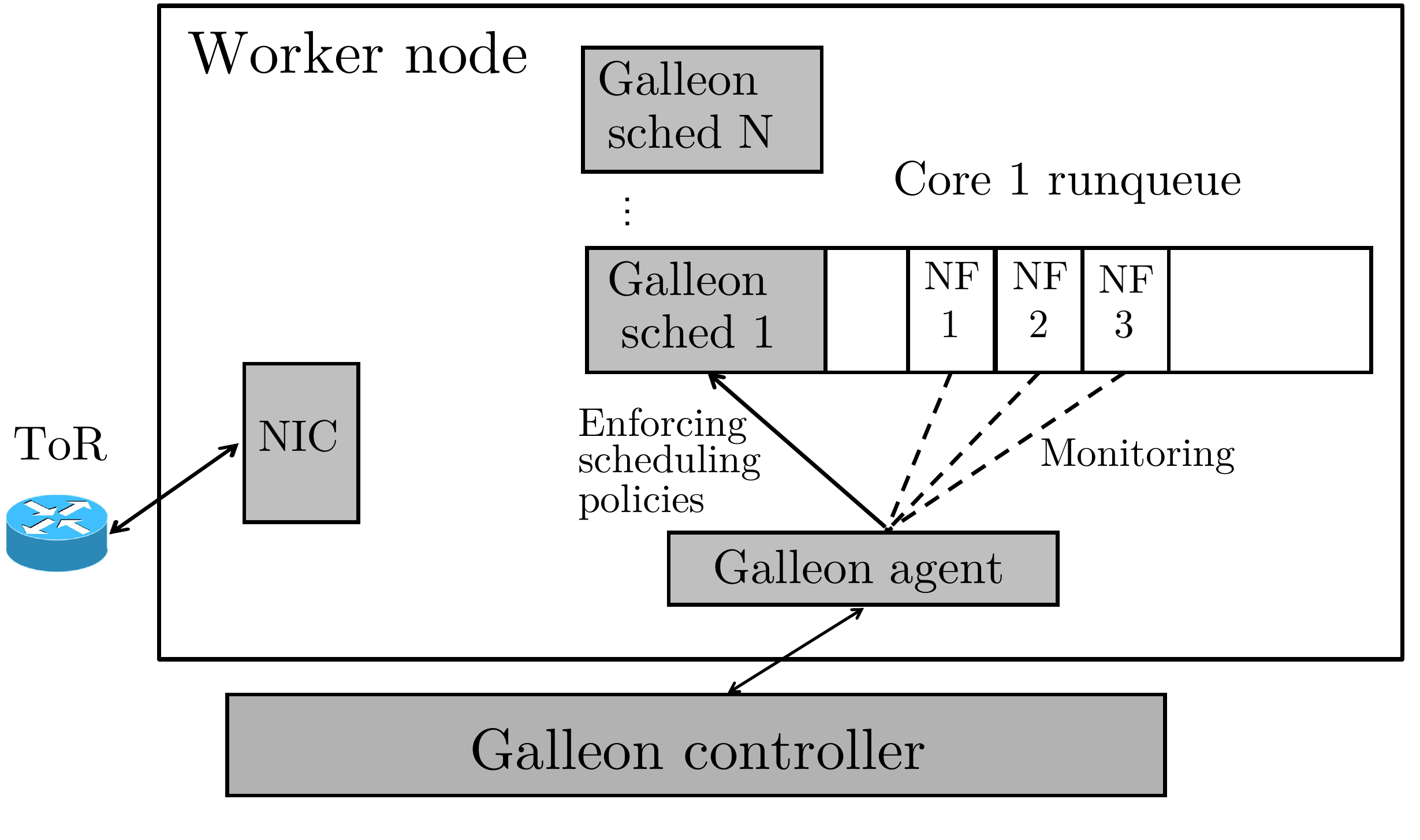}
}
}
\caption{\sysname reuses a FaaS worker subsystem to run both system components and function instances. \sysname controller interacts with \sysname ingress and FaaS worker subsystem to deploy NFs in serving packet processing functions. Unshaded boxes are existing FaaS components, lightly shaded ones are FaaS components that \sysname modifies,  and darker ones represent new components specific to \sysname hidden within the infrastructure.}
\label{fig:quadrant-arch}
\end{center}
\vskip -1em
\end{figure}

Figure~\ref{fig:quadrant-arch} illustrates the key \sysname control-plane components. \sysname reuses the basic FaaS architecture, making (a) minimal modifications to the FaaS interface, and (b) augmenting existing FaaS components with mechanisms, instrumentation, and algorithms to support fast packet processing.

\subsection{\sysname Interface}
\label{s:user-interface}

\sysname must provide a general abstraction for writing user-defined NFs. Most FaaS platforms today employ a RESTful API to handle network events, usually via HTTP. They allow developers to write a customized function as the event handler that takes an event struct as input, parsed from a HTTP request’s payload.
This abstraction is not ideal for processing raw packets. Unlike dedicated requests to the FaaS’s ingress server (via a public IP), \sysname must permit each packet to be processed by one or more NFs. To do this without modifying FaaS’s programming model, one may, at the ingress,  encapsulate each packet within a HTTP request payload. This can introduce significant extra overhead~\cite{ServelessNF}.

\sysname modifies the FaaS programming and deployment model minimally to support packet processing. NF developers can accept a raw packet struct (a pointer to the packet) as input. A \sysname operator (\eg an ISP) can assemble an NF chain using these NFs. The customer then submits to \sysname the NF chain, a traffic filter specification for what is to be processed by the chain, and a per-packet latency SLO.

\subsection{\sysname Design}

\sysname extends FaaS functionality as shown in \figref{fig:quadrant-arch}. It assumes 
commodity servers and OpenFlow-enabled switches. Components that are unique in \sysname, hidden within the infrastructure, include: the \sysname controller, the \sysname NF runtime, the \sysname ingress, the \sysname scheduler, and a per-worker \sysname agent that acts as a representative of the centralized \sysname controller.

\sysname re-uses the existing FaaS worker subsystem for managing workers and deploying services. The worker subsystem (implemented using Kubernetes in many FaaS implementations~\cite{openfaas}) manages the system resources, \ie a pool of worker machines. Each worker machine executes functions encapsulated in containers; in \sysname, each container hosts an NF.  Customers provide container images for each NF in a chain. The centralized \sysname controller manages the deployment of FaaS services by interacting with this worker subsystem to deploy \sysname components (an ingress, per-worker scheduler and agents) prior to startup.

At runtime, the \sysname controller uses the worker subsystem to deploy NFs as containers. It collects NF performance statistics reported from each agent, serves queries from the ingress to make scheduling decisions, and pushes load balancing decisions to the ingress which enforces it by modifying a ToR's flow table. The \sysname ingress is where NFV traffic enters and leaves the system. It works with an Openflow-enabled switch to enforce the \sysname controller's workload assignment strategies. When a new flow arrives at the switch, packets are forwarded to the \sysname ingress that determines the target logical NF chain for this flow. To do so, it queries the \sysname controller to pick a chain for this flow, then applies the \sysname controller's load balancing for this logical chain to pick a deployed chain. Then it installs a flow rule to offload the flow dispatching task to the switch hardware. This design reduces the traffic sent to the software ingress module, and thus decreases CPU time spent routing packets.

In such a setup, a ToR switch only passes customer-specified aggregates to \sysname's ingress. Traffic that needs application-level processing will reach a normal FaaS ingress service that schedules requests as a normal FaaS system does. In this way, \sysname coexists with normal FaaS processing and does not impact its performance.

\section{\sysname's Runtime}
\label{s:sysnames-nf-runtime}

\sysname's main challenge is FaaS compatibility. Next we describe the design of \sysname's runtime to achieve performance, scaling, and isolation through careful allocation of NIC and CPU resources, and by careful CPU scheduling.

\subsection{NF Execution Models}
\label{s:design-space-nf}

Prior work has explored different NF execution models that dictate how NFs share packet memory, how the runtimes steer packets to NFs, and how they schedule NF execution.

\parab{Memory model.} A FaaS instance may reside in memory for some time before it times out. During its lifetime, packets and NF-internal state are loaded into memory for processing. For NFs running on the same worker machine, they (1) may share NF state memory, and packet buffer memory (e.g., in Metron~\cite{Metron2018} and NetBricks~\cite{NetBricks2016}), (2) do not share NF state memory, but share packet buffer memory globally (\eg in E2~\cite{E22015} and NFVNice~\cite{kulkarni2017nfvnice}), or (3) do not share either NF state memory or packet buffer memory (\eg in EdgeOS~\cite{edgeOS2020}).

\parab{Network I/O model.} Packets must be sent to a specific NF running on a specific server core. In many NFV platforms, such as E2~\cite{E22015}, NFVNice~\cite{kulkarni2017nfvnice}, and EdgeOS~\cite{edgeOS2020}, a hardware switch (presumably with Openflow) forwards packets to specific worker machines. Once packets arrive at the server’s NIC, a virtual switch forwards traffic locally. In a multi-tenant environment, the vSwitch has read and write access to each individual NF’s memory space, and copies packets when forwarding them from an upstream NF to its downstream. The vSwitch is a core service and can become the bottleneck for both intra- and inter-machine traffic. To scale it up, a runtime can add CPU cores for vSwitches, but cores can generate revenue, so this strategy is undesirable. On our test machine, a CPU core can only achieve a 6.9~Gbps throughput when forwarding 64-byte packets (or 13.5~Mpps). Consider a chain with 4 NFs running on a server with a 10~Gbps NIC. The aggregate traffic volume can reach 40~Gbps at peak, which requires at least $7$ CPU cores to run vSwitches (more if traffic is not evenly distributed across the vSwitches).

An alternative approach is to offload packet switching to the ToR switch and the NIC’s internal switch. Both switches coordinate to ensure packets arrive at the target machine’s/target process’s memory. When a packet hits the ToR, the switch not only forwards the packet to a dedicated machine, but also facilitates intra-machine forwarding via L2 tagging. This approach eliminates the need to run a vSwitch. However, it can only ensure that packets are received by the first NF in a chain. Metron and NetBricks take this approach, but rely on a strong assumption: that all NFs can be compiled and run in a single process. However, many popular NFs are only available in a closed-source form from commercial software developers, and they cannot be compiled with other NFs to form a single binary that runs the NF chain. Even if that were possible, the packet isolation requirement constrains flexibility significantly, since it can only then be achieved using language-level memory isolation (\eg by using Rust).

\parab{CPU scheduling model.} Memory and network I/O models also impact CPU resource allocation and scheduling of NFs and NF chains. When NF chains run in a single process (as in Metron and NetBricks), those runtimes can dedicate a core to an entire chain. When NFs run in separate processes (as in E2 or NFVnice), runtimes must decide whether to allocate one or more cores to a chain, and how to schedule each NF.

\subsection{\sysname's Execution Model}
\label{s:nf-runtime}

To ensure minimal changes to existing FaaS architectures, \sysname chooses an execution model that sits in a different point in the design space. \sysname deploys each NF in an NF chain as a single container. From the perspective of the underlying FaaS platform, an NF chain represents a ``function'' (a unit of invocation). NFs can share packet buffers (as in Metron or NetBricks), but packet isolation is enforced through OS protection, careful scheduling and packet copying (unlike NetBricks which relies on language level isolation). \sysname uses NIC I/O virtualization and kernel bypass to reduce packet steering overhead. The rest of the section describes some of the details of \sysname's packet I/O and memory models. The next section describes CPU allocation and scheduling.

\parab{Packet I/O.} \sysname uses DPDK for fast userspace networking to handle packet I/O for NFs. However, because \sysname must co-exist with other FaaS functions that may be executing on the same CPU, it must use userspace networking only for NF chains. To do so, \sysname uses the hardware virtualization technique Single-root Input/Output Virtualization\cite{sriov-intro} (SR-IOV) to virtualize the NIC hardware.\footnote{SR-IOV is a PCIe specification that allows a PCIe device to appear as many physical devices, providing a hardware method for sharing a hardware resource with isolation. With SR-IOV,  NIC hardware generates one Physical Function (PF) that controls the physical device, and many Virtual Functions (VFs) that are lightweight PCIe functions with the necessary hardware access for receiving and transmitting data.} On a worker machine, the \sysname agent (\figref{fig:quadrant-arch}) manages the virtualized devices via kernel APIs through the PF. Once the \sysname controller allocates CPU core to an NF chain, the \sysname agent sets up a VF to steer NF-chain traffic to that core.

\parab{Memory.}
In \sysname, a runtime on behalf of an NF chain initializes a file-backed dedicated memory region that holds fixed-size packet structures for incoming packets. It also creates a ring buffer that holds packet descriptors that point to these packet structures. To receive packets from the virtualized NIC, the NF runtime passes this ring buffer to its associated VF so that the NIC hardware can perform DMA directly to the NF runtime’s memory.

\parab{NF State Management.} For stateful NFs, packet processing depends on both the packet itself and the NF’s current internal state. Prior work (\eg statelessNF~\cite{kablan2017stateless}, S6~\cite{S6}, SNF~\cite{SNF}) has proposed maintaining NF global state in an external store (such as Redis~\cite{sanfilippo2009redis});  they adopt various caching schemes to mitigate the latency overhead of pulling state from the external store. In \sysname, we adopt a similar scheme by maintaining per-NF global state remotely and maintaining general NF state in a flow table locally at each NF runtime instance. At the front-end, the programming model provides a set of simple APIs for writing a stateful NF: \texttt{\small update(flow, val)} and \texttt{\small read(flow, val)}, where flow corresponds to a BPF matching rule. The NF runtime makes the process transparent and interacts with the external store to pull state when necessary and synchronize state periodically. \sysname can incorporate prior efforts on optimizing state access latency because the state management model remains similar and the optimizations do not require modified APIs.

\subsection{Core Allocation and Scheduling}
\label{sec:cooperative-sched}

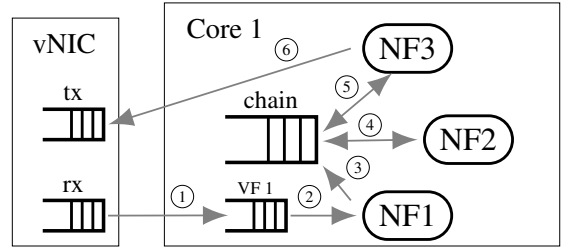
\begin{figure}[t]
\begin{center}
\resizebox{\columnwidth}{!}{%
  \tikzset{%
  box/.style = {
    draw,
    align = center,
    inner sep = 4pt,
    text centered,
  },
  arrow/.style = {
    ->,
    gray,
    text = black,
    line width = .25pt,
  },
}

\newcommand*\circled[1]{%
  \tikz[baseline=(char.base)]{%
    \node[shape=circle,draw,ultra thin,inner sep=2pt,scale=.225] (char) {#1};}}

\def\queue {
  \draw[yshift=.5cm,line width=.5pt,align=center]
  (0,0) -- ++(2cm,0) -- ++(0,-1cm) -- ++(-2cm,0);
  \foreach \Val in {1,...,3}
  \draw ([yshift=.5cm,xshift=-\Val*10pt]2cm,0) -- ++(0,-1cm);
}

\begin{tikzpicture}[scale=.3,>=latex]

  \draw[ultra thin,box] (0,0) rectangle (1.75,3.75);
  \node[above,scale=.4] at (.85,3) {vNIC};

  \begin{scope}[shift={(.5,.5)},scale=.5]
    \node[above,scale=.35] at (1cm,.5) (rxq) {rx};
    \queue
  \end{scope}
  \begin{scope}[shift={(.5,2)},scale=.5]
    \node[above,scale=.35] at (1cm,.5) (txq) {tx};
    \queue
  \end{scope}

  \node[above,scale=.4] at (3.5,3.25) {Core 1};
  \draw[ultra thin,box] (2.5,0) rectangle (9,4);

  \begin{scope}[shift={(3.5,1.75)},scale=.75]
    \node[above,scale=.35] at (1cm,.5) (chq) {chain};
    \queue
  \end{scope}

  \begin{scope}[shift={(3.5,.5)},scale=.5]
    \node[above,scale=.25] at (1cm,.5) (vfq) {VF 1};
    \queue
  \end{scope}

  \node[box,rounded rectangle,scale=.45] at (6.5, .5) (nf1) {NF1};
  \node[box,rounded rectangle,scale=.45] at (7.5, 1.75) (nf2) {NF2};
  \node[box,rounded rectangle,scale=.45] at (6.5, 3.25) (nf3) {NF3};

  \draw[arrow] ([shift={(.575,-.25)}] rxq.south) -- node [above,midway,yshift=-2.5,xshift=2] {\circled{1}} ([shift={(-.05,-.45)}] vfq.west);

  \draw[arrow] ([shift={(.575,-.25)}] vfq.south) -- node [above,midway,yshift=-2.5,xshift=-2] {\circled{2}} ([shift={(-.1,0)}] nf1.west);

  \draw[arrow] ([shift={(-.2,.25)}] nf1.west) -- node [above,midway,yshift=-3,xshift=3] {\circled{3}} ([shift={(.85,-.8)}] chq.south);

    \draw[arrow, <->] ([shift={(.85,-.4)}] chq.south) -- node [above,midway,yshift=-3] {\circled{4}} ([shift={(-.1,0)}] nf2.west);

    \draw[arrow, <->] ([shift={(.85,-.25)}] chq.south) -- node [above,midway,yshift=-3,xshift=-1.5] {\circled{5}} ([shift={(-.25,0)}] nf3.south);

  \draw[arrow] ([shift={(-.2,0)}] nf3.west) -- node [above,midway,xshift=8] {\circled{6}} ([shift={(.575,-.25)}] txq.south);

\end{tikzpicture}
}
\caption{Timeline of packets on a \sysname worker. A packet is tagged at the ingress. 1) NIC's L2 switch sends it to NIC VF associated with the destined chain. NIC VF DMAs packets to the first NF's memory space. 2) NF 1 processes the packet. 3) After NF 1's packet processing function returns, the packet is copied to the chain's pktbuf by the NF runtime if there are other NFs. This is necessary to ensure packet isolation as the NIC's pktbuf should only been seen by NF 1. 3)-5) A per-core cooperative scheduler controls the execution sequence of NFs to ensure temporal packet isolation. 6) Final NF asks VF to send the packet out.} %
\label{fig:quadrant-packet-timeline}
\end{center}
\vskip -1em
\end{figure}

Userspace I/O and shared memory can reduce overhead significantly, but to be able to process packets at high throughput and low latency, \sysname must have tight control over NF chain execution. As discussed earlier, customized NF platforms use two different approaches. One approach bundles NFs in an NF chain into a single process to \textit{run to completion} (in which each NF in the chain processes a batch of packets before moving onto the next batch (as in Metron or NetBricks). This approach ensures performance predictability and high performance by amortizing overhead over a packet batch. To achieve packet isolation, NetBricks relies on language isolation. The second approach, which NFVNice uses, is to run each NF in a separate process, which ensures isolation with copying packets, but can require careful allocation of CPU shares, and orchestration of process execution on the underlying scheduler (\eg CFS in NFVNice).

In contrast to these two approaches, we introduce \textit{spatiotemporal packet isolation} in which NF chains operate on 1) spatially isolated packet memory regions (as opposed to the typical model in run-to-completion software switches such as BESS, in which all NF chains on a machine run in the same memory) and 2) are temporally isolated through careful sequencing of their execution, which proceeds in a run-to-completion fashion \emph{across processes} and uses \textit{cooperative scheduling} mechanisms to hand off control and the natural execution boundary of packet batch handoff.  This isolation ensures that NF chains (which may process different customers' traffic) cannot see each others packet streams or state, and even within a chain each NF maintains private state and only gets to execute (and thus access packet memory) when it is expected to perform packet processing in the chain.

In \sysname, NFs are individual containers deployed in a Kubernetes cluster. \sysname dedicates a core to a chain that actively serves traffic. This choice is appropriate for our setting, since we expect that NF chains will handle large traffic aggregates, so that multiple cores might be necessary to process traffic for a given chain (\secref{sec:faas-controller}).  The \sysname controller manages all NFs via Kubernetes APIs to control the allocation of memory, CPU share, and disk space.

\subsubsection{Enforcing run-to-completion scheduling}
\label{sec:enforcing-scheduling}

\sysname relies on a per-core cooperative scheduler to schedule NFs on each core. All NF containers that are part of a chain are assigned to a single core. Each NF container runs an NF Runtime in a multi-process way for executing an user-defined NF. The NF process runs in a single-thread for processing traffic. The other process is transparent to NF authors. It runs a RPC server for controlling the NF's behaviour, and a monitoring thread that collects performance statistics.

Threads are transparent to NF authors, and are handled by \sysname completely. NF authors just develop a standard NF with \sysname's interface. \sysname's runtime takes the NF code and executes it in the NF process, while the monitoring thread (\secref{sec:faas-controller}) collects performance statistics. A \sysname agent in the NF runtime communicates these statistics to the \sysname controller, and handles NF chain instantiation as well as NF-chain scaling (\secref{sec:faas-controller}). To avoid interfering with packet processing, the monitoring thread runs on a separate core.

To tightly coordinate NF chain execution, \sysname uses Linux's real-time (RT) scheduling support, and manages NF threads real-time priorities and schedules them using a FIFO policy. We use this policy to emulate, as described below, NF chain run-to-completion execution in which each NF in the chain processes a batch of packets in sequence.

\parab{Scheduling model.} In \sysname's cooperative scheduling, an upstream NF runs in a loop to process individual packets of a given batch, and then yields the core to its downstream NF. This is transparent to the NFs: once the user-defined NF finishes processing, it invokes the NF runtime to transmit the batch  to the downstream NF; the runtime invokes yield.\footnote{To deal with non-responsive NFs, the runtime terminates chain execution if an NF fails to yield after a conservative timeout.}

For this, the cooperative scheduler has to bypass the underlying scheduler (CFS in our implementation) and take full control of a core. Internally, the scheduler maintains two FIFO queues: a run queue that contains runnable NF processes, and a wait queue that contains all idle NF processes. It offers a set of APIs that the NF runtime can use to transfer the ownership of NF processes of a chain from CFS to the cooperative scheduler.
These APIs are used by the \sysname agent, which runs as a privileged process. NF threads are not aware of these APIs and cannot interact with coopsched to change scheduling policy or core affinity.

Once a chain is deployed, all NF threads are managed by the Cooperative Scheduler, and are placed in the scheduler's wait queue as \emph{detached}. Once an NF chain switches into the \emph{attached} state, the Cooperative Scheduler pushes NFs of this chain into its run queue and ensures that the original NF dependencies are preserved in the run queue. To detach a chain, the Cooperative Scheduler waits for the chain to finish processing a batch of packets, if any, and then moves these NF processes back to the wait queue.

\parab{How scheduling works.}
The \sysname controller deploys multiple NF runtime instances that contain NF processes in a chain. Once an NF process starts, \sysname's NF runtime reports its thread ID (tid) to the \sysname agent running on the same worker. Once all NF processes are ready, the \sysname agent \textit{registers} them as a scheduling group (sgroup) to the Cooperative Scheduler. After that, the cooperative scheduler takes full control of NF processes.

When the \sysname controller assigns flows to an NF chain (\secref{sec:faas-controller}), the Cooperative Scheduler \textit{attaches} the chain to the core. When the monitoring thread sees no traffic has arrived for the chain, the Scheduler \textit{detaches} the chain, so the \sysname controller can re-assign the core.

To effect attach and detach operations, and to schedule NF chain execution, the Cooperative Scheduler has a master thread for serving scheduling requests and runs one enforcer thread on each managed core. The scheduler utilizes a key feature of Linux FIFO thread scheduling: 1) high-priority threads preempt low-priority threads; 2) a thread is executed once it is at the head of run queue, and is moved to the tail after it finishes. An enforcer thread is raised to the highest priority when enforcing scheduling decisions. When an NF chain is instantiated on a core, the enforcer thread registers the corresponding NF processes as low-priority FIFO threads so that they are appended to the wait queue. When attaching the NF chain, it moves NF processes to the run queue by assigning them a higher priority, and vice versa when detaching a sgroup. Operations are done in the sequence that NFs are positioned in the NF chain, so when an NF yields, the CPU scheduler automatically schedules the next NF in the chain.

In this model, each worker machine splits CPU cores into two groups. One group is managed by the Cooperative Scheduler, while the other runs with normal threads managed by CFS. NF processes run in a normal during it startup time.
We use a standard kernel and support different schedulers on different cores. This enables running running NF and non-NF workloads on the same machine.

\parab{Adaptive batch size.} The Cooperative Scheduler introduces \(N\) context switches for a chain with \(N\) NFs. Without an appropriate batch size, the core may spend a significant portion of time on these context switches. \sysname applies adaptive batch scheduling to bound the context switch overhead within a threshold of total CPU execution time. This optimization largely applies to long but lightweight chains. In some cases, deploying such a lightweight long chain with a low latency requirement prevents us from achieving large batch sizes at execution time. However, a lightweight chain is often executed with high per-core throughput. Such chains easily serve tens of Gbps of traffic with just a few cores. As such, the absolute number of extra CPU cores remains small.

For an NF chain $v$ with $N$ NFs, suppose $B_{v} \in \mathbb{N}$ is the max packet batches that the first NF should process before it sends it downstream. We estimate the packet rate $R_{v}$ that $v$ can achieve at runtime as follows: $Freq$ is the CPU core's clock rate. $T_{v,i}$ is the $i$th NF's service time in CPU cycles, $T_{ctx}$ is the profiled time duration of a context switch in seconds. $b_{v}$ is the average number of packets per batch at the input of a chain. $b_{v}$ depends on both the traffic arrival pattern and the chain service pattern, and is limited by the hardware, i.e. $b_{v}\in[1,b_{m}]$ where $b_m \in \mathbb{N}$ is the maximum number of packets in a batch.

\begin{equation}
R_{v} = \frac{Freq}{\sum_{i=1}^{N}T_{v,i} + N \cdot T_{ctx} \cdot Freq \cdot \frac{1}{B_v \cdot b_{v}}}
\end{equation}

Suppose $\hat{R_{v}}$ is the packet rate when running $v$ in a single thread. The goal is to bound the downgraded maximum per-core packet rate within a ratio $p$ of the maximum rate $\hat{R_{v}}$. The batch size for chain $v$ is given by:
\begin{equation}
min\ B_v
\end{equation}
s.t.
\begin{equation}
R_{v} >= p \cdot \hat{R_{v}}, \quad where \quad\hat{R_{v}} = \frac{Freq}{\sum_{i=1}^{N}T_{v,i}}
\end{equation}
Simple algebraic manipulations suffice to compute $B_v$.

\subsection{Spatiotemporal Packet Isolation}
\label{s:inter-nf-comm}

Current FaaS offerings constrain functions to use distributed storage for transmitting data between functions. This can significantly increase latency of packet processing; \eg Redis client to server communication, even when the client and server are on the same rack, incurs an average latency of 310~$\mu$s, and a worst-case latency of 47~ms. Across multiple NFs in a chain, the resulting latency can be substantial.

\sysname can use standard IPC mechanisms or shared memory to communicate between NFs. To permit zero-copy transfer, \sysname uses shared memory, but its design must ensure packet isolation (\secref{sec:problem-definition}); a chain should not be able to see packets belonging to another chain, and an NF should not be able to access a packet before its upstream NF has processed it or after it has completed its processing.

Figure~\ref{fig:quadrant-packet-timeline} describes how \sysname achieves this packet isolation. Each distinct NF chain is allocated a separate virtual NIC queue with SR-IOV; moreover, different NF chains run in different containers. This ensures spatial packet isolation across NF chains. Ensuring packet isolation between NFs within a single chain is harder. Ideally, the virtual NIC could directly DMA packets to shared memory so all NFs can process packets without copies. However, this would violate isolation because a downstream NF could access shared memory while the NIC writes to it.

To avoid this, \sysname gives only the first NF in the chain access to the NIC queue, to process a batch of packets and copy them to shared memory. Thus, the first NF's packet memory is \textit{spatially} isolated from other chains and from downstream NFs. Cooperative scheduling ensures NFs run in the order they appear in the chain, so \textit{temporal} packet isolation is also preserved: even though it has access to shared memory, a downstream NF cannot access a batch that has not been processed by an upstream NF since it will not be scheduled. At the same time, this permits zero-copy packet transfer between all NFs except the first.

For a chain with only one NF, \sysname omits the unnecessary packet copying and cooperative scheduling. The \sysname NF runtime also applies an optimization that prefetches packet headers into the L1 cache before calling the user-defined NF for processing tasks. As shown in \ref{subsec:executing-nf-chain}, this optimization improves performance by 11\% when running a canonical NF.

\section{Auto-scaling in \sysname}
\label{sec:faas-controller}

\sysname \textit{auto-scales} (automatically adapts resources allocated to) NF chains in response to changes in traffic volume. In a canonical FaaS platform~\cite{openfaas,wang2018peeking}, a FaaS controller  interacts with a global ingress service and worker machines to coordinate auto-scaling. The global ingress service maintains a reception queue for incoming requests. The FaaS controller maintains a group of function instances on the workers. For each user-defined function, as new requests arrive, the ingress service dispatches each request to an idle instance. The FaaS controller scales up and down the number of instances to handle dynamic traffic while achieving cost efficiency.

\sysname employs a similar architecture (\figref{fig:quadrant-arch}), modified to satisfy performance and isolation goals. Unlike canonical FaaS workloads, NFV workloads typically have an SLO that specifies the target latency~\cite{tootoonchian2018resq,YenWang2020Lemur}. \sysname's controller is designed to serve dynamic traffic while meeting stringent latency SLOs. The controller performs two functions: 1) balance load across the workers; 2) create/delete new instances of NF runtimes, and schedule/un-schedule NF threads among the workers. In this, it is aided by a per-worker \sysname agent that monitors NF's performance and interacts with the cooperative scheduler to enforce scheduling policies. The rest of the section describes components of the controller and the algorithms that it uses for auto scaling.

\subsection{Monitoring}
\label{s:monitoring}

Monitoring is critical for scaling NF chains. At each NF runtime, an NF monitoring function collects performance statistics, including the instantaneous packet rate, NIC queue length, and the per-batch execution time. The packet rate is measured as the average processing rate of the NF chain. NIC queue length is the number of packets reported by the NIC hardware. It also estimates per-batch execution time by recording the global CPU cycle counter at the beginning and the end of sampled executions. %

To avoid interfering with data-plane processing, the NF monitoring function runs in a separate thread and is not scheduled on a core running NF processing threads. Each NF runtime maintains statistics locally, and sends updates to the \sysname controller only when significant events occur (to minimize control overhead). Examples of such  events include queue lengths or rates exceeding a threshold.

\subsection{\sysname ingress}
\label{subsec:system-ingress}

In a typical \sysname deployment (\eg at an ISP edge), the traffic aggregate hitting an NF chain is likely to exceed the capacity of a single core. In general, then, multiple instances of an NF chain might be deployed across the workers. The load balancer component interacts with the \sysname ingress to split flows among the deployed NF chains.

Consider a rack deployment of \sysname: all worker machines are on a rack and connected to a ToR switch.  External traffic enters the rack at this ToR switch. When a new flow hits the ToR switch, it does not match any rules, so the ToR forwards the flow to \sysname's ingress (a software component running on one of the workers). The ingress then queries the \sysname controller to select an NF chain to assign to this flow. It then installs a flow entry on the ToR switch to route subsequent packets of this flow to the worker running the NF chain instance. To do this, \sysname relies on L2 tagging to  forward the flow to the worker's virtual NIC. Prior work~\cite{SNF} proposes to make such assignments at a flowlet (sub-flow) level to better balance load in the face of different flow sizes. \sysname can incorporate this optimization, which we have left this to future work.

\sysname could have performed this reactive rule installation using an OpenFlow controller. We found that this approach resulted in packet losses caused by high flow-entry-installation latency. Our ingress is implemented as a service and can buffer packets until the flow entry is installed, thereby avoiding packet loss.

The \sysname controller makes load balancing decisions based on statistics received from each worker (\secref{s:monitoring}). For each flow, it always assigns the NF chain with the highest traffic load (the current packet rate / the maximum packet rate) among all non-overloaded chains. To determine a load threshold for setting a chain as overloaded, \sysname applies a profiling-based solution to pick the highest threshold that results in a latency bounded by the SLO requirement.

\subsection{Scaling of NF Chains}
\label{subsec:nf-chain-scaling-algorithm}

\sysname attempts to schedule NF chains on the fewest number of CPU cores that can serve traffic without violating SLOs. It uses a combination of mechanisms to achieve this.

First, to avoid cold start latency and the resulting packet loss, \sysname maintains a pool of pre-deployed NF chains; it leverages prior work for reducing the startup time of serverless instances~\cite{oakes2018sock,manco2017lightvm}. Pre-deployed NF threads start out in the detached state, so they do not consume CPU resources. %

Second, \sysname profiles the NF chain to determine the maximum packet rate achievable on a single core. When the traffic assigned to an NF chain approaches that rate, \sysname decides to allocate new incoming flows to a new NF chain instance. This is accomplished by the \sysname ingress (\secref{subsec:system-ingress}).

Finally, when an NF thread becomes idle (all flows previously assigned to it have completed), \sysname reclaims the assigned core. \sysname could have, instead, migrated flows away from underutilized NF chains, but this would have complicated state management for stateful NFs. We have left this optimization to future work.

Listing \ref{lst:scaling-nf-chains-in-cluster} describes the algorithms used at the ingress to assign flows to NF chain instances, and the algorithms used at the worker's cooperative scheduler to dynamically attach and detach cores from NF chains in response to traffic dynamics. Together, these algorithms achieve auto scaling in \sysname.

\begin{lstlisting}[float, style=PyStyle, caption=Pseudocode for scaling NF chains in a \sysname cluster, label=lst:scaling-nf-chains-in-cluster]
def handle_flows_at_ingress(flow,
                            chains: NFChainsList):
 # Each logical chain is associated with 
 # a list of active NF chains v.sgroups, 
 # and a list of idle NF chains v.idle_sgroups
 for v in chains:
  if v.should_serve(flow):
   selected=v.pick_highest_load_chain_without_overload()
   if selected is None:
    selected=v.pick_idle_chain_from_lowest_load_worker()
   selected.set_active()
   update_routing(match=flow,
                  action=(set_port=selected.port,
                          set_l2=selected.l2_tag))
   while len(v.idle_sgroups) < scale_out_thresh:
    scale_out(v)
   while len(v.idle_sgroups) > scale_in_thresh:
    scale_in(v)

def worker_scheduler_loop(worker: GalleonWorker):
 while True:
  for sg in worker.sgroups:
   if sg.is_sched and not sg.is_active:
    worker.detach_sgroup(sg)
   # sg is set active by the above ingress
   if sg.is_active and not sg.is_sched:
    idle_core=worker.pick_idle_core()
    worker.attach_sgroup_to_core(sg, idle_core)
\end{lstlisting}
\vskip -1em

\section{Implementation}
\label{sec:implementation}

\sysname is built upon OpenFaaS~\cite{openfaas}.  OpenFaaS  consists of an infrastructure layer and an application layer.
The former uses three components: Kubernetes, Docker, and the Container Registry. \sysname reuses these APIs for managing and deploying NFs. At the application layer, OpenFaaS has its own gateway service for triggering function executions. \sysname adds a separate ingress component. Incoming traffic is split at the system gateway; requests for normal applications are forwarded to OpenFaaS's gateway, while NFV traffic is forwarded to the \sysname ingress. OpenFaaS uses a function runtime that maintains a tunnel to the FaaS gateway, and hands off requests to user-defined functions. \sysname uses a different NF runtime by using the above mechanisms for receiving traffic from the \sysname ingress (\secref{sec:faas-controller}). \sysname reuses OpenFaaS's general monitoring framework, but relies on a per-worker agent for NF performance monitoring and enforcing scheduling policies.

\parab{\sysname controller.} This multi-threaded controller is implemented in Go and adds 9K LOC to OpenFaaS. It controls the \sysname ingress and the worker cluster subsystem via RPCs and Kubernetes APIs. On each worker, this controller deploys an agent that runs as a RPC server for collecting NF performance statistics, updating critical performance events, and enforcing scheduling policies. The controller also includes a UI subsystem. Operators may use this UI system to deploy NF chains, adjust cooperative scheduling and batch scheduling policies, and monitor NF chains by clearly seeing which chain runs at which core in the cluster with its instantaneous packet rate, load and NIC queue length.

\parab{\sysname ingress.} This is implemented in 800 LOC of C++ as a BESS~\cite{bess} module that receives \sysname controller's load balancing decision and enforces it by modifying the hardware switch table via OpenFlow.

\parab{\sysname Cooperative Scheduler.} This is implemented in C as a multi-threaded process in 5K LOC. It offers a set of APIs that can be used by agents.

\parab{\sysname NF Runtime.} This is implemented in C++ by reusing some BESS libraries (with 2.5K LOC including an RPC server and a driver that works with \sysname's network I/O). It runs two processes: one for RPC server and one for NF. The NF runtime uses an external datastore (i.e. a Redis cluster) for maintaining global NF states.

\section{Evaluation}
\label{sec:evaluation}

\label{subsec:eval-methodology}

\parab{Experiment setup.} We use a Cloudlab~\cite{cloudlab-paper} cluster of 10 servers, each with dual-CPU 16-core 2.4GHz Intel Xeon(R) CPU E5-2630 v3 (Haswell) with 128GB ECC memory (DDR4 1866 MHz). To reduce jitter, all CPU cores have hyperthreading and CPU frequency scaling disabled and run at a fixed 2.4 GHz. Each server has one dual-port 10~GbE Intel X520-DA2 NIC. Both are connected to an experimental LAN for data-plane traffic. Each machine has one 1~GbE Intel NIC for control and management traffic. Servers connect to a Cisco C3172PQs ToR switch with 48 10~GbE ports and Openflow v1.3 support. The traffic generator and the \sysname ingress run on dedicated machines.

\parab{Methodology and Metrics.} Our \sysname experiments employ end-to-end traffic. We deploy three representative chains from light to heavy in their CPU cycle cost, from documented use cases~\cite{ietf-sfc-use-cases-06}. Chain 1 implements a lightweight L2/L3 pipeline for tunneling: Tunnel$\rightarrow$IPv4Forward; Chain 2 implements a complex chain with DPI and encryption NFs: ACL$\rightarrow$UrlFilter$\rightarrow$Encrypt; Chain 3 is a state-heavy chain that requires connection consistency: ACL$\rightarrow$NAT.
Key performance metrics include: the end-to-end latency distribution and packet loss rate, and the time-average and maximum CPU core usage for the test duration. The traffic generator uses BESS to generate flows with synthetic test traffic.

\subsection{Executing NF chains}
\label{subsec:executing-nf-chain}

We explore how \sysname's performance changes when running an NF chain of various lengths. We compare the per-core max throughput achieved by \sysname and other NFV systems that make different isolation choices. For simplicity, we use a test NF (a BPF~\cite{linux_bpf} module with a table of 200 BPF rules) and run chains with a sequence of the same NF.

\parab{Isolation via copying.} EdgeOS~\cite{edgeOS2020} supports isolation using data copying.  We emulate EdgeOS on top of a reimplementation of NFVNice~\cite{kulkarni2017nfvnice} and apply the same packet copying technique to the master module that acts as the message bus for transmitting packets between processes. The master module runs as a multi-threaded process with one RX thread for receiving packets from the NIC, one TX thread for transmitting packets among processes, and one wake-up thread for notifying a process that a message has arrived at its message buffer. All three threads run on dedicated cores.

\parab{Isolation via safe languages.} NetBricks~\cite{NetBricks2016} uses compile-time language support from Rust to ensure isolation among NFs plus a run-time array bound check.

\begin{figure}[t]
  \resizebox{\columnwidth}{!}{\pgfplotsset{
  /pgfplots/ybar legend/.style={
    /pgfplots/legend image code/.code={%
      \draw[##1,/tikz/.cd,yshift=-0.25em]
      (0cm,0cm) rectangle (5pt,0.75em);},
  },
  Base/.style = {
    fill,
  },
  Quadrant/.style = {
    Base,
    red,
  },
  Metron/.style = {
    Base,
    orange,
  },
  NetBricks/.style = {
    Base,
    lightgray!95!blue,
  },
  NFVNice/.style = {
    Base,
    cyan,
  },
}

\begin{tikzpicture}[scale=1]

  \pgfplotstableread{figs/data/isolation_comp_tput.dat}\chainlendata;

  \begin{axis}
   [
    font = \large,
    width = .75\textwidth,
    height = .35\textwidth,
    xlabel = {NF Chain Length},
    ylabel = {Throughput [Mpps]},
    tick pos = left,
    major tick length = 2,
    minor tick length = 1.5,
    xlabel near ticks,
    ylabel near ticks,
    xtick align = outside,
    ytick align = outside,
    xtick = {1,2, ..., 7},
    ytick = {0,0.5, ..., 4},
    minor ytick = {0, 0.25,...,4},
    ybar,
    legend cell align = left,
    legend style = {
      font = \normalsize,
      row sep = 1.5pt,
    },
   ]
   \addplot [Quadrant] table [x expr=\thisrowno{0}, y expr=\thisrowno{2}/1000] {\chainlendata};
   \addplot [Metron] table [x expr=\thisrowno{0}, y expr=\thisrowno{1}/1000] {\chainlendata};
   \addplot [NetBricks] table [x expr=\thisrowno{0}, y expr=\thisrowno{3}/1000] {\chainlendata};
   \addplot [NFVNice] table [x expr=\thisrowno{0}, y expr=\thisrowno{4}/1000] {\chainlendata};

   \addlegendimage{Quadrant};
   \addlegendentry{\sysname};
   \addlegendimage{Metron};
   \addlegendentry{\sysname (single thread)};
   \addlegendimage{NetBricks};
   \addlegendentry{NetBricks};
   \addlegendimage{NFVNice};
   \addlegendentry{NFVNice w/ pkt copy};

  \end{axis}
\end{tikzpicture}}
\caption{Throughput with increasing chain length for running an NF chain on a single core when using 80-byte packets.}
\label{fig:isolation_comp_tput}
\end{figure}
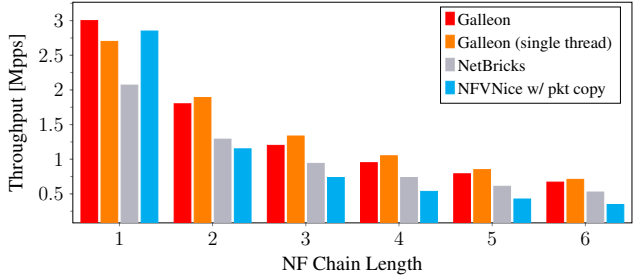

\parab{Results.}
Figure~\ref{fig:isolation_comp_tput} shows the throughput of different isolation approaches for different length chains. \sysname outperforms both NetBricks (1.28-1.44$\times$ perf) and NFVNice w/ packet copying (1.05-1.93$\times$ perf). NFVNice with packet copying has similar performance (95\%) with a single-NF chain, but as chain length increases its throughput decreases despite its 3 extra CPU cores for transmitting packets among NFs. 

We also implemented a variant labeled \sysname (single thread) that runs all NFs in a single thread. This variant offers no isolation because all NFs run in the same process, with neither language support for isolation nor our spatiotemporal packet isolation mechanism. Compared to this unsafe-but-fast variant, \sysname has an overhead that remains at the same level regardless of the chain length: \sysname achieves a 89.9\text{\%}-95.2\text{\%} per-core throughput when deploying a multi-NF chain while providing isolation. For a chain with one NF, \sysname achieves slightly better performance as it applies the prefetch-into-L1 optimization described earlier.

\subsection{Scaling Experiments}
\label{subsec:eval-compresults}

Next we evaluate \sysname's performance when deploying NF chains with latency SLOs.

\subsubsection{Controlling NF chain latency}
\label{subsec:control-nf-chain-latency}

Prior to the deployment of an NF chain, \sysname profiles the chain to get its latency profile under different per-core load thresholds. When deploying a chain, the \sysname controller considers the user-specified NF chain latency SLO requirement, and uses the per-core load threshold as a knob to control the end-to-end delay for packets being processed by NF chains deployed in the cluster.

\begin{figure}[t]
    \centering
    \resizebox{\columnwidth}{!}{\pgfplotsset{
  ProfileGroupPlot/.style = {
    group/group size = 1 by 2,
    group/xlabels at = edge bottom,
    group/xticklabels at = edge bottom,
    group/ylabels at = edge left,
    group/vertical sep = 6pt,
  },
  ProfilePlot/.style = {
    width = .6\textwidth,
    height = .25\textwidth,
    xlabel = {Per Core Load Threshold [\%]},
    tick pos = left,
    major tick length = 2,
    minor tick length = 1,
    xlabel near ticks,
    ylabel near ticks,
    xmin = -3,
    xmax = 90,
    xtick align = outside,
    xtick = {2,5,10,15, ..., 85},
  },
  Line/.style = {
      thin,
      dotted,
      mark size = 2pt,
      forget plot,
    },
    LatencyLine/.style = {
      Line,
      color=red,
      mark = *,
    },
    QueueLine/.style = {
      Line,
      color = olive,
      mark = triangle*,
    },
    RateBar/.style = {
      forget plot,
      ybar,
      fill,
      color = brown,
      bar width = 6pt,
      nodes near coords,
      every node near coord/.append style={
        anchor=south,
        font=\fontsize{4}{6},
        white!15!black,
      },
    },
}
\pgfplotstableread{figs/data/chain1_profile.dat}\profiledata
\begin{tikzpicture}[scale=1]
  \begin{groupplot}[ProfileGroupPlot]

    \nextgroupplot[
    ProfilePlot,
    xlabel = {},
    ylabel = {Rate [kpps]},
    ymin = 75,
    ymax = 650,
    ytick = {100, 200, ..., 600},
    minor ytick = {100, 125, ..., 650},
    ]
    \addplot [RateBar] table [x expr=\thisrowno{0}, y expr=\thisrowno{1}] {\profiledata};

    \nextgroupplot[
    ProfilePlot,
    ylabel = {Tail Latency [$\mu{}s$]},
    ymin = 0,
    ymax = 180,
    ytick = {40, 70, ..., 170},
    minor ytick = {40, 50, ..., 162},
    legend style = {
      font = \small,
      at = {(.35,.925), anchor = west},
    },
    ]
    \addplot [LatencyLine] table [x expr=\thisrowno{0}, y expr=\thisrowno{3}/1000] {\profiledata};
    \addplot [QueueLine] table [x expr=\thisrowno{0}, y expr=\thisrowno{2}] {\profiledata};

   \addlegendimage{LatencyLine};
   \addlegendentry{Tail Latency};
   \addlegendimage{QueueLine};
   \addlegendentry{Queue Length};

  \end{groupplot}

  \begin{groupplot}[ProfileGroupPlot]
    \nextgroupplot[
    ProfilePlot,
    xlabel = {},
    ylabel = {},
    ymin = 75,
    ymax = 650,
    ]

    \nextgroupplot[
    ProfilePlot,
    ytick pos = right,
    yticklabel pos = right,
    ylabel = {Queue Len [packets]},
    ymin = 0,
    ymax = 180,
    ytick = {10, 40, 70, 100, 128},
    minor ytick = {10, 15, ..., 128},
    ]
  \end{groupplot}

\end{tikzpicture}}
    \caption{NF Chain Profiling: Effect of per-core load threshold on tail latency (p99), queue length, and rate.}
    \label{fig:nf_chain_profile}
    \vskip -0.5em
\end{figure}
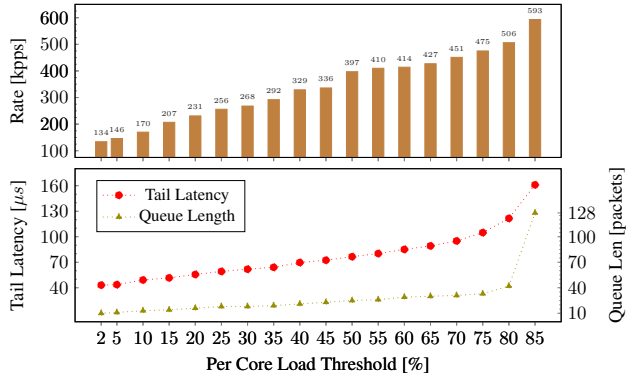

\parab{Results.}
Figure~\ref{fig:nf_chain_profile} shows profiling results for NF chains used in the cluster-scale experiments. As shown in the figure, the NF chain's tail latency is reduced when the \sysname controller gradually decreases the per-core load threshold when running a chain. According to the profiling results, we see the cause of this behaviour: by reducing the per-core load threshold for a chain, the chain runs at an operating point where the maximum queue length is lower. This is because the \sysname controller considers a chain as overloaded if its current traffic load exceeds the threshold, and stops assigning new flows to this chain. Therefore, this design limits the amount of traffic destined to a chain, and further controls the maximum queue length observed by the chain, and therefore controls the chain's end-to-end latency.

This feature, of course, aligns with the trade-off between latency and efficiency: a smaller tail latency means a smaller per-core throughput. When serving the same amount of traffic, it means that the \sysname controller has to devote more CPU cores for running this chain. This unique feature is also very important in the FaaS context: \sysname is able to use the right level of system resources to meet the latency SLO when deploying NFV chains.

\subsubsection{The end-to-end evaluation}
\label{subsec:end-to-end-eval}

We compare \sysname against a customized, high-performance NFV platform, Metron~\cite{Metron2018}. Like \sysname, Metron also utilizes the hardware switch for dispatching traffic. Metron compiles NFs into a single process, and runs-to-completion each chain as a thread. We have implemented both Metron's runtime and scaling algorithm. Each Metron runtime is a single-thread process that runs an NF chain. Packets are transmitted between NFs with no isolation mechanisms (as Metron does not consider isolation).

We compare \sysname and Metron in these end-to-end experiments. Before experiments, an NF chain specification is passed to both system's controllers to deploy NFs in the test cluster. For each experiment, we run a DPDK-based flow generator to generate traffic at 20 flows/s with a median packet size of 1024-byte. We estimate this level of flow rate by analyzing an ISP trace~\cite{caida}. All incoming flows are forwarded to the system ingress and to be processed by their destined NF chains. The traffic generator gradually increases the number of flows and reaches the maximum throughput after 60 seconds, and results in an average of 1.2K flows at the peak load of 18~Gbps. This number of flows is similar to that used by prior work~\cite{kablan2017stateless,SNF}. Then the traffic generator stays at the steady state to keep generating traffic at the maximum rate until the total number of packets reaches the 100 million. We compare end-to-end performance metrics, including the tail latency, the maximum number of cores, and the time-averaged number of cores for deploying the chain.

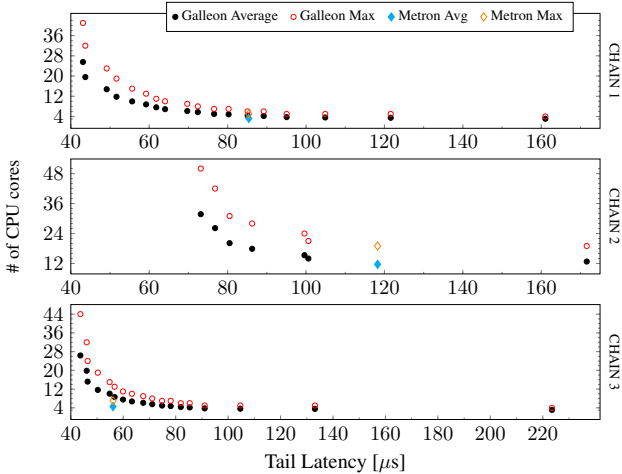
\begin{figure}[t]
    \centering
    \resizebox{\columnwidth}{!}{\pgfplotsset{
  /pgfplots/ybar legend/.style={
    /pgfplots/legend image code/.code={%
      \draw[##1,/tikz/.cd,yshift=-0.25em]
      (0cm,0cm) rectangle (3pt,.6em);},
  },
  CPUUtilGroupPlot/.style = {
    group/group size = 1 by 3,
    group/xlabels at = edge bottom,
    group/ylabels at = edge left,
    group/vertical sep = 15pt,
    xlabel = {Tail Latency [$\mu$s]},
    xlabel near ticks,
    ylabel near ticks,
  },
  CPUUtilPlot/.style = {
    legend columns = 4,
    legend cell align = left,
    legend style = {
      font = \scriptsize,
       at = {(.95,1.1), anchor = south},
      /tikz/every even column/.append style={column sep=0.15cm},
    },
    width = .6\textwidth,
    height = .2\textwidth,
    major tick length = 2,
    minor tick length = 1,
    xtick pos = bottom,
    xticklabel style={
            /pgf/number format/fixed,
            /pgf/number format/precision=0,
            /pgf/number format/fixed zerofill
        },
  },
  Marker/.style = {
    only marks,
    mark = *,
    forget plot,
  },
  Avg/.style = {
    black,
    Marker,
    mark size = 1.25pt,
  },
  Max/.style = {
    red,
    Marker,
    mark size = 1.25pt,
    mark = o,
  },
  Metron/.style = {
    cyan,
    Marker,
    mark=diamond*,
    mark size = 2pt,
  },
  MetronMax/.style = {
    orange,
    Marker,
    mark=diamond,
    mark size = 2pt,
  },
}

\pgfplotstableread{figs/data/cpu_chain1.dat}\chainIdata
\pgfplotstableread{figs/data/cpu_chain2.dat}\chainIIdata
\pgfplotstableread{figs/data/cpu_chain3.dat}\chainIIIdata

\begin{tikzpicture}[scale=1]
  \begin{groupplot}[CPUUtilGroupPlot]
  \nextgroupplot[
    CPUUtilPlot,
    ytick = {4, 12, ..., 41},
    minor ytick = {3, 4, 6, ..., 42},
    xtick={40,60, ..., 160},
    minor xtick={40, 50,..., 170},
    xmin=40,
    xmax=175,
   ]
   \addplot[Avg] table [x expr=\thisrowno{1}/1000, y=AvgCores] \chainIdata;
   \addplot[Max] table [x expr=\thisrowno{1}/1000, y=MaxCores] \chainIdata;
   \addplot[Metron] coordinates {(85.4, 3.22)};
   \addplot[MetronMax] coordinates {(85.4, 5)};

    \addlegendimage{Avg};
    \addlegendentry{\sysname{} Average};
    \addlegendimage{Max};
    \addlegendentry{\sysname{} Max};
    \addlegendimage{Metron};
    \addlegendentry{Metron Avg};

    \addlegendimage{MetronMax};
    \addlegendentry{Metron Max};

  \nextgroupplot[
    CPUUtilPlot,
    ytick = {12, 24, ..., 50},
    minor ytick = {12, 14, ..., 50},
    xtick={40,60, ..., 160},
    minor xtick={40, 50,..., 170},
    xmin=40,
    xmax=175,
   ]
   \addplot[Avg] table [x expr=\thisrowno{1}/1000, y=AvgCores] \chainIIdata;
   \addplot[Max] table [x expr=\thisrowno{1}/1000, y=MaxCores] \chainIIdata;
   \addplot[Metron] coordinates {(118.25, 11.66)};
   \addplot[MetronMax] coordinates {(118.25, 19)};

  \nextgroupplot[
    CPUUtilPlot,
    ytick = {4, 12, ..., 44},
    minor ytick = {3,4, 6, ..., 44},
    xtick={40,60, ..., 230},
    minor xtick={40, 50,..., 230},
    xmin=40,
   ]
   \addplot[Avg] table [x expr=\thisrowno{1}/1000, y=AvgCores] \chainIIIdata;
   \addplot[Max] table [x expr=\thisrowno{1}/1000, y=MaxCores] \chainIIIdata;
   \addplot[Metron] coordinates {(56.1, 4.54)};
   \addplot[MetronMax] coordinates {(56.1, 7)};

 \end{groupplot}

  \node[xshift=-1cm] at
  ($(group c1r1.west)!0.5!(group c1r3.west)$)
  {\rotatebox{90}{{\small \# of CPU cores}}};

  \foreach \i in {1,...,3}
  \node[xshift=.25cm] at
  ($(group c1r\i.north east)!0.5!(group c1r\i.south east)$)
  {\rotatebox{270}{{\scriptsize CHAIN \i}}};

\end{tikzpicture}}
    \caption{Core usage of NF chains implemented in \sysname and Metron as a function of achieved tail latency. Per-core thresholds are set as in \figref{fig:nf_chain_profile}.}
    \label{fig:cpu_util}
    \vskip -1em
\end{figure}

\parab{Results.}
Across all experiments, both systems achieve a zero loss rate. Thus, we compare two systems by looking at the tail (p99) latency of the CPU core usage when they serve the test traffic (100 million packets). Figure~\ref{fig:cpu_util} shows this for different latency targets. For all chains, \sysname can precisely meet the SLO for its tail latency. Metron ignores the latency SLO and targets a deployment without losses.

Take NF chain 1 as an example. Metron fails to meet a low latency SLO target (50 us). Both \sysname and Metron can satisfy the medium latency SLO (100 us). In terms of CPU usage, we notice that \sysname and Metron use the same number of cores at peak. \sysname uses a slightly higher time-averaged number of cores (15.5\%). We see this gap as the overhead of achieving isolation among NFs.

\subsection{Overhead}
\label{subsec:eval-microbench}

Next, we show a breakdown of \sysname overhead. \sysname's overhead comes from spatiotemporal isolation, including 1) the spatial isolation of a virtual NIC to isolate chains and packet copy from the NIC's buffer to the NF chain's buffer, and 2) the temporal isolation provided by transparent cooperative scheduling, with extra context switches among NFs (unlike in a single process).

\subsubsection{Spatial isolation overhead}
\label{subsec:spatial_isolation_overhead}

To evaluate \sysname's spatial isolation overhead, we quantify the SR-IOV NIC's overhead by comparing performance metrics when running a test NF with and without enabling SR-IOV. Our test NF is a MACSwap module that swaps the dst and src Ethernet addresses of a packet.

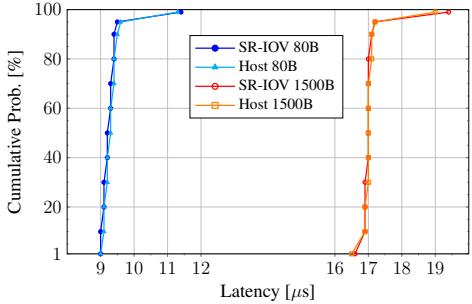
\begin{figure}[t]
  \centering
  \resizebox{.75\columnwidth}{!}{
    \centering
    \begin{tikzpicture}

  \pgfplotstableread{figs/data/sriov_comp_delay.dat}\delaymeas;

  \begin{axis}[
    height = .4\textwidth,
    width = .6\textwidth,
    EcdfLine/.style = {
      draw,
      thick,
      mark size=1.25pt,
    },
    VirtLine/.style = {
      EcdfLine,
      mark = *,
    },
    HostLine/.style = {
      EcdfLine,
      mark = triangle*
    },
    font = \large,
    xtick pos=bottom,
    xmajorgrids,
    ymajorgrids,
    major tick length = 2,
    minor tick length = 1,
    ytick={1,20,40,...,100},
    minor ytick = {0,5,...,100},
    ymin=1,
    ymax=100,
    xmin = 8,
    xmax = 20,
    xtick = {9,10,11,12, 16,17,18,19},
    minor xtick = {9,9.5, ..., 12, 16,16.5,...,19.5},
    xtick align=outside,
    ylabel={Cumulative Prob. [\%]},
    xlabel={Latency [$\mu$s]},
    legend image post style={scale=1.5},
    legend cell align = left,
    legend style={
      font=\normalsize,
      at={(.305,.725)},
      anchor=west},
    ]
    \addplot[VirtLine,blue] table [y=p, x expr = \thisrowno{1} / 1000] {\delaymeas};
    \addlegendentry{SR-IOV 80B};
    \addplot[HostLine,cyan] table [y=p, x expr = \thisrowno{2} / 1000] {\delaymeas};
    \addlegendentry{Host 80B};
    \addplot[VirtLine,red,mark=o] table [y=p, x expr = \thisrowno{3} / 1000] {\delaymeas};
    \addlegendentry{SR-IOV 1500B};
    \addplot[HostLine,orange,mark=square] table [y=p, x expr = \thisrowno{4} / 1000] {\delaymeas};
    \addlegendentry{Host 1500B};
  \end{axis}
\end{tikzpicture}
  }%
\caption{End-to-end latency CDF with SR-IOV on and off.}
\label{fig:sriov-comp}
\end{figure}

\parab{Results.}
Figure~\ref{fig:sriov-comp} shows that running the NF with an SR-IOV enabled NIC only increases the latency by 0.1 us for both 80-byte and 1500-byte packets. We also find that the maximum throughput achieved by a SR-IOV enabled NIC is always greater than or equal to 99.6\% of the throughput achieved by a NIC running in a non-virtualized mode.

\subsubsection{Temporal isolation overhead}
\label{subsec:temporal_isolation_overhead}

For a chain, the ownership of a packet is transferred between NFs as the chain proceeds. Packet isolation requires that NFs in the same chain should only acquire the ownership of a packet after its predecessor finishes processing it (\secref{s:inter-nf-comm}). To quantify this temporal isolation overhead, we use a multi-NF chain in which the first NF holds a NIC VF that is dedicated to this chain. The NIC hardware DMAs incoming packets to the NIC packet buffer that resides in this NF's memory space. The packet isolation requirement prevents other NFs from accessing the NIC packet buffer directly, as packets in that memory region are only destined to the first NF. For the rest of the chain, NFs share the access of a per-chain packet buffer and wait to be scheduled by the Cooperative Scheduler to process the same batch of packets in the correct sequence.

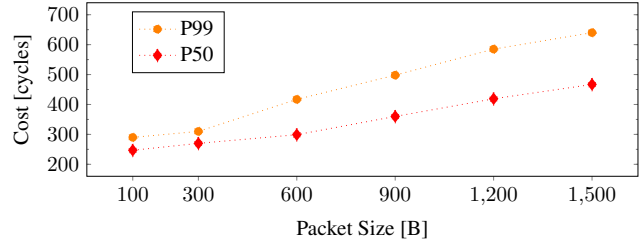
\begin{figure}[t]
  \resizebox{\columnwidth}{!}{\begin{tikzpicture}[scale=1]

  \pgfplotstableread{figs/data/copying_packet_overhead.dat}\copydata;

  \begin{axis}
   [
    width = .6\textwidth,
    height = .25\textwidth,
    xlabel = {Packet Size [B]},
    ylabel = {Cost [cycles]},
    xtick pos = bottom,
    major tick length = 2,
    minor tick length = 1,
    xlabel near ticks,
    ylabel near ticks,
    ymin = 160,
    ymax = 740,
    xtick align = outside,
    xtick = {100,300, 600, ..., 1500},
    ytick = {200, 300, ..., 700},
    minor ytick = {200, 250, ..., 700},
    legend style = {
      at = {(.25,.95), anchor = west},
    },
    Line/.style = {
      mark = *,
      thin,
      dotted,
      mark size = 2pt,
      forget plot,
    },
    PVLine/.style = {
      Line,
      red,
      mark = diamond*,
      mark size = 3pt,
    },
    PIXLine/.style = {
      Line,
      orange,
    },
    ]

   \addplot [PVLine] table [x expr=\thisrowno{0}, y expr=\thisrowno{1}] {\copydata};
   \addplot [PIXLine] table [x expr=\thisrowno{0}, y expr=\thisrowno{2}] {\copydata};

   \addlegendimage{PIXLine};
   \addlegendentry{P99};
   \addlegendimage{PVLine};
   \addlegendentry{P50};

  \end{axis}
\end{tikzpicture}}
\caption{Per-packet cost of copying packets of different sizes} 
\label{fig:copying_packet_overhead}
\end{figure}

\parab{Results.}
Figure~\ref{fig:copying_packet_overhead} shows the p50 and p99 CPU cycle cost for copying one packet of different sizes. At median, copying a 100-byte packet costs 247 cycles, while doing so for a 1500-byte packet costs 467 cycles. The small difference of copying a small and a large packet is because the cost of allocating a packet struct takes a significant share in the total cost.

Scheduling NFs cooperatively involves context switches between NF threads that belong to different NF processes. We profile the average cost of context switches between NFs: 2143 cycles per context switch. Note that this context switch cost is amortized among the batch of packets in each execution. For a default 32-packet batch, the amortized cost is only 67 cycles per packet. This cost is 27 $|$ 14\% of the cost for copying a 64 $|$ 1500-byte packet respectively. Further, it is only 31\% of the cost for forwarding a packet via a vSwitch with packet copying, as in NFVNice. (\sysname has zero packet switching cost because it uses the ToR switch and the NIC's L2 to dispatch packets to different chains.)

\parab{Alternative packet ownership transfer.}
We also consider a traditional mechanism for ensuring isolation for a shared memory region in a multi-process environment by using \texttt{\small unmap} and \texttt{\small map} to explicitly move the ownership of the shared packet buffer. We profile the average cost for these two options:  \texttt{\small unmap} is 4083 cycles, while \texttt{\small map} is 8495 cycles. With all packets placed in the same memory page, we need one \texttt{\small unmap} and \texttt{\small map} to transfer the memory page to a different process. We conclude that their costs are significantly larger (5.87$\times$) than the context switch overhead.

\parab{Cooperative scheduling effects.}
For a chain, cooperative scheduling involves context switches between different NF processes.
This operation can also flush caches and TLBs; we conduct an experiment to understand these effects.

Here we run the same test chain, with 5 BPF modules, as in \ref{subsec:executing-nf-chain} with four experimental groups: 1) \sysname: the vanilla deployment w/o adaptive batch optimization; 2) Local mem: the vanilla deployment that operates on one dummy packet in the shared memory region; 3) Dummy NF: a chain of dummy NFs that do not process packets, but simulate NF cycle costs; 4) Single thread: a chain that runs in a single thread. None use adaptive batch optimization so that they run with the same batch size. One traffic generator produces excessive traffic (1024B packets) to saturate the chain's NIC queue so that each chain runs at a batch size of 32, the NIC's default batch size. The TLB and cache misses are measured as the average value for a 15-second execution duration for 5 measurements.

\parae{Results.}
Table~\ref{t:nf_execution_performance} shows NF runtime stats. For all multiple-process groups, we see higher iTLB and dTLB misses. As shown, the number of dTLB misses reaches less than 1\% of dTLB hits for cases that run a non-dummy NF, though dTLB misses are less important for an NF's performance.

All multi-process groups see higher iTLB misses compared to the single thread case because NF processes do not share code in memory. The `local mem' and `dummy NF' cases perform similarly in terms of per-packet cycle cost (and the number of cache misses). This is because the `local mem' case processes one packet that resides in the chain's local memory and is likely to benefit from the L3 cache. The `\sysname' case has a slightly higher per-packet cost compared to the other two multi-process cases. Breaking down per-chain cycle cost into the per-NF level, we find that the extra cost only comes from the first NF that copies incoming packets. The 2nd-4th NFs in 1)-3) have the same cycle cost (509 cycles / packet). These NFs benefit from L3 caching as the first NF's runtime loads when copying packets from the NIC's packet buffer to the per-chain packet buffer.

In the above four cases, two major differences explain the per-chain cycle cost: a) iTLB misses when deploying as a multi-process chain and b) L3 cache misses when processing network traffic. The `\sysname' case has both; the `Local mem' and `Dummy NF' cases have a); and `Single thread' has b). We can calculate the breakdown cycle cost for each via simple algebraic manipulations for this 5-NF chain: a) 254 cycles / packet (or 50.8 cycles / packet / hop); b) 100 cycles / packet. Among these two, a) is the extra overhead of a context switch, which could be reduced by tagged TLBs; b) is not an overhead as a chain must load packets into the CPU cache once when processing network traffic. Thus the amortized TLB overhead is relatively small compared to the context switch itself.

\begin{table}[t]
\centering
\scriptsize
\resizebox{\columnwidth}{!}{
\begin{tabular}{p{.5cm}rrrrr}
  \multicolumn{2}{c}{\textbf{Metric}} & \textbf{\sysname} & \textbf{Local mem} & \textbf{Dummy NF} & \textbf{Single thread} \\[1pt]
  \toprule
  \multicolumn{2}{l}{Per packet cycles} & 2846 & 2746 & 2730 & 2592 \\[3pt]
  \multirow{2}{*}{\shortstack[l]{Chain delay\\ (in cycles)}} & P50 & 121536 & 116232 & 116008 & 85980 \\
  & P99 & 128116 & 122324 & 117892 & 87492 \\[3pt]
  \multirow{5}{*}{Misses} & dTLB & 72,864,218 & 68,259,827 & 61,251,661 & 1,185,591 \\
  & dTLB (\%) & 0.52\% & 0.48\% & 4.07\% & 0.00\% \\
  & iTLB & 11,542,564 & 12,325,256 & 9,615,381 & 591,737 \\
  & iTLB (\%) & 478.18\% & 488.28\% & 379.31\% & 182.39\% \\
  & Cache & 18,923,855 & 6,710,255 & 6,699,896 & 12,963,766 \\
  & L1 dcache & 508,578,460 & 417,298,551 & 333,262,806 & 327,837,034 \\
  & L1 icache & 41,272,281 & 37,127,027 & 30,856,178 & 17,568,605 \\
  \bottomrule
\end{tabular}
}
\caption{Overheads under isolation variants.}
\label{t:nf_execution_performance}
\end{table}

\section{Related Work}
\label{sec:related}

\parab{Serverless platforms.}
Serverless platforms are available as both commercial cloud services\cite{AWSLambda2017,ApacheOpenWhisk,AzureFunctions,GoogleCloudFunctions}, and as open-source offering~\cite{openfaas,OpenLambda2016}. They are designed for latency-insensitive applications, and not suited for running NFV workloads for reasons that have been discussed along this paper.

Research on serverless platforms has explored two issues. The first improves aspects of FaaS platforms. Sock~\cite{oakes2018sock} and LightVM~\cite{manco2017lightvm} optimizes sandbox startup time. SAND~\cite{akkus2018sand} optimizes IPC performance. E3~\cite{liu2019e3} accelerates serverless execution with SmartNIC to improve cost-efficiency. The second explores new applications only made possible with serverless, such as real-time big data analysis~\cite{227653,klimovic2018pocket} and scalable video encoding~\cite{fouladi2017ExCamera}. ServerlessNF~\cite{ServelessNF,SNF} executes NFs on serverless platforms. While offering insights on handling different flow sizes, it does not propose changes to the infrastructure, and thus does not leverage techniques for building high-performance NFV platforms. Indeed, its main focus is on deploying a single NF without meeting SLOs for NF chains~\cite{tootoonchian2018resq, YenWang2020Lemur}. ServerlessNF's centralized resource manager (RM) incurs extra overhead of CPU cores and NIC bandwidth, and poses a new challenge to make RM scalable while meeting SLOs.

\parab{NFV frameworks.} NFV frameworks~\cite{E22015,Metron2018,NetBricks2016,YenWang2020Lemur,kulkarni2017nfvnice} deploy, execute, and orchestrate NF chains in a highly optimized way. E2~\cite{E22015} deploys NF chains using commodity servers, but lacks optimizations and isolation mechanisms. Subsequent work explored a number of specific topics. \textbf{Isolation:} NetBricks~\cite{NetBricks2016} isolates NFs with a safe runtime based on Rust. This assumes that NF vendors offer source code written in a certain language, which is uncommon in practice. In contrast, most commercial NFs are packaged containers or VMs without source. EdgeOS~\cite{edgeOS2020} employs an expensive isolation mechanism via packet copying for each NF. \textbf{Hardware acceleration:} a number of papers examine accelerating NFV by leveraging specialized hardware such as P4 switches~\cite{UNO2017,Metron2018,YenWang2020Lemur}. \sysname sees the NFV's fundamental motivation as reducing infrastructure management cost, and does not use specialized hardware platforms that are not widely  available in today's cloud infrastructure. \textbf{Performance tuning:} Some~\cite{tootoonchian2018resq,kulkarni2017nfvnice,batchy} examine dirty-slate solutions to optimize NFV's performance. They often work in a restricted setting without considering multi-tenant cluster deployment; that said, this line of work is largely compatible with \sysname and could be integrated for further performance gains.

\parab{Stateful NFs.} Another line of research~\cite{S6,kablan2017stateless,SNF} considers optimizations for stateful NFs. \sysname supports stateful NFs but does not specifically implement some of these more advanced techniques; these are complementary mechanisms that can be integrated into \sysname.

\section{Conclusions}
\label{sec:conclusions}

\sysname supports NFV in edge and cloud computing environments using commodity software and hardware components, fulfilling NFV's original ambitions. It minimally extends FaaS abstractions, and eases the deployment of third-party NFs. With its spatiotemporal packet isolation mechanism, it outperforms state-of-the-art NFV platforms that use alternative isolation mechanisms, and has performance comparable to custom NFV systems that do not consider NF isolation.

\begin{small}

  \bibliography{references.bib}
\end{small}

\end{document}